# Graphene promotes axon elongation through local stall of Nerve Growth Factor signaling endosomes

D. Convertino[1,2,*], F. Fabbri.[2,†], N. Mishra[2], M. Mainardi[4,‡], V. Cappello[2], G. Testa[4], S. Capsoni[4], L. Albertazzi[5,6], S. Luin[1,3], L. Marchetti[2,7,*,#], C. Coletti[2,*,#]

[1] NEST, Scuola Normale Superiore, Pisa, Italy

[2] Center for Nanotechnology Innovation @NEST, Istituto Italiano di Tecnologia, Pisa, Italy

[3] NEST Istituto Nanoscienze—CNR and Scuola Normale Superiore, Pisa, Italy

[4] BIO@SNS Laboratory, Scuola Normale Superiore, Pisa, Italy

[5] Nanoscopy for Nanomedicine Group, Institute for Bioengineering of Catalonia (IBEC), The Barcelona Institute of Science and Technology (BIST), Carrer Baldiri Reixac 15-21, 08024 Barcelona, Spain

[6] Department of Biomedical Engineering, Institute for Complex Molecular Systems (ICMS), Eindhoven University of Technology, 5612AZ Eindhoven, The Netherlands

[7] Department of Pharmacy, University of Pisa, Pisa, Italy

**Corresponding Author**
[*] domenica.convertino@iit.it; laura.marchetti@unipi.it; camilla.coletti@iit.it;

[#] joint last co-authors

**Present Addresses**
[†] NEST Istituto Nanoscienze—CNR and Scuola Normale Superiore, Pisa, Italy

[‡] IN-CNR Institute of Neuroscience, Consiglio Nazionale delle Ricerche, Pisa Italy

**ABSTRACT**

Several works reported increased differentiation of neuronal cells grown on graphene; however, the molecular mechanism driving axon elongation on this material has remained elusive. Here, we study the axonal transport of nerve growth factor (NGF), the neurotrophin supporting development of peripheral neurons, as a key player in the time course of axonal elongation of dorsal root ganglion neurons on graphene. We find that graphene drastically reduces the number of retrogradely transported NGF vesicles in favor of a stalled population in the first two days of culture, in which the boost of axon elongation is observed. This correlates with a mutual charge redistribution, observed via Raman spectroscopy and electrophysiological recordings. Furthermore, ultrastructural analysis indicates a reduced microtubule distance and an elongated axonal topology. Thus, both electrophysiological and structural effects can account for graphene action on neuron development. Unraveling the molecular players underneath this interplay may open new avenues for axon regeneration applications.

**INTRODUCTION**

In the past years, a number of studies have investigated graphene potential as a conductive neural interface, able to enhance adhesion, proliferation and differentiation of various cell types, including neural cells [1–4]. In the last case, an interesting increased neurite sprouting and outgrowth induced by graphene were reported for hippocampal neurons [1], differentiated SH-SY5Y neuroblastoma cell line [5], adrenal phaeochromocytoma (PC12) cell line [4,6,7], dorsal root ganglion (DRG) neurons [4] and neural stem cells [8]. Such feature makes graphene appealing for application in peripheral nerve regeneration, where an appropriate scaffold may accelerate neurite outgrowth [9,10]. However, to date, few studies have examined the interaction of graphene with peripheral neurons. Thus, the nanoscale mechanisms by which graphene would promote axon regeneration remain unclear.

To understand the molecular mechanism of axon outgrowth on graphene, some groups investigated GAP-43, a recognized marker of developing and regenerating axons. Increased GAP-43 levels were reported in hippocampal neurons grown on graphene [1]. Similarly, increased GAP-43 together with synaptophysin levels were also reported in PC12 cells grown on aligned silk-graphene hybrid hydrogels [6], suggesting that the mechanisms driving neuritogenesis may be shared for central and peripheral neurons grown on graphene. However, the underlying mechanisms for the increased GAP-43 expression on graphene, together with structure and dynamics of neuron development have not yet been investigated. A recent study has reported an increased cell firing for hippocampal neurons grown on graphene, probably due to altered membrane ion currents at the material interface [11]. Indeed, several reports described a positive effect of electrical stimulation on axonal outgrowth and branching [3,12–14]. However, such effect may not necessarily apply to peripheral neurons on graphene, as it was already reported that DRG neurons respond differently from hippocampal neurons on nanofabricated biomaterial scaffolds [15,16]. Importantly, all the above-cited studies were carried out on cultures grown from one to three weeks onto the substrate. While this strategy is useful to understand the long-term effects of the material on neuron physiology, it fails to investigate how the early developmental stage is influenced by the neuron-material interface.

This appears to be an important issue to address, because the positive effect of graphene on axonal sprouting and outgrowth was found to be maximal during the first two days of culture, and then to decrease up to a steady-state level in which axons are slightly longer with respect to control cultures [1].

In this work we demonstrate that graphene significantly stimulates axonal outgrowth during the first 3 days in culture of DRG primary neurons. Hence, we investigate the effect of graphene on the fast axonal transport properties during the early developmental phase. In detail, the role of nerve growth factor (NGF), a key neurotrophin involved in neurite elongation and survival during DRG development [17,18] is investigated by single-molecule fluorescence microscopy. Remarkably, we find that axonal elongation on graphene correlates with a significant reduction of NGF vesicles retrogradely transported to the soma, in favor of a stalled pool retained locally in developing axons. Patch-clamp recordings and ultrastructural analyses concur to show that profound rearrangements occur in axons developing on graphene, which may account for the local accumulation of NGF resulting in the increased axon elongation. Our results provide a broad structural and functional understanding of the impact of graphene on DRG neurons, key information towards the development of graphene-based devices for neural regeneration applications.

**RESULTS AND DISCUSSION**

**Graphene promotes axon elongation in developing DRG neurons**

The effect of graphene on axon outgrowth in primary peripheral neurons was investigated using DRG neurons dissected from postnatal day (P) 4 mice, which were cultured on PDL/laminin coated substrates [4] in the presence of NGF. In order to precisely determine axonal length, we used a compartmentalized microfluidic chamber placed on top of a graphene-coated glass coverslip (Figure 1a).

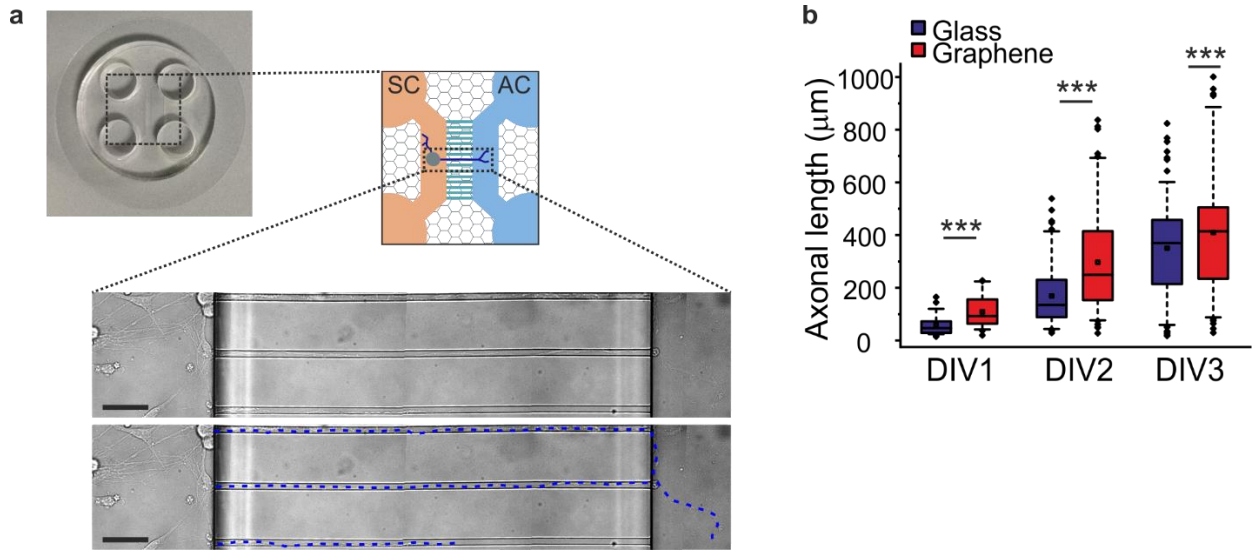

*Figure 1. Increased axonal elongation in DRG neurons grown on graphene (a) Schematic representation of DRG axonal length quantified on graphene inside a microfluidic chamber. Top left, optical image of an assembled device on graphene. Top right, cartoon showing the microfluidic chamber with the soma compartment (SC) and axon compartment (AC) connected by microchannels. Bottom: bright field images with axons crossing and filling the microchannels. The same area is reported below with the axons highlighted by dashed blue lines. Scale bars: 50 μm. (b) Quantification of the axonal length at different DIV on glass and graphene in a microfluidic device. A significant difference was found between graphene and control according to Two-way Anova Test. (P_substrate <0.001, Two-way Anova, with Holm-Sidak post-hoc test, \*\*\*p<0.001 ). The number of measured axons was: DIV1, glass n=42, graphene n=50; DIV2, glass n=109, graphene n=101; DIV3, glass n=147, graphene n=119. In box plots: box between 25th and 75th percentile; horizontal line: median; whiskers: $5^{th}$ and $95^{th}$ percentiles; square: mean; circles: outliers.*

This approach allows to selectively apply treatments to axons and mimics well the physiological NGF stimulation – where the trophic factor is produced by target innervated tissues and is endocyted at the distal axon tips, far from the cell soma [19,2]>.

Figure 1b reports the axon length distribution measured at three time points. We found gradually increasing neurite lengths both on graphene and glass, with significantly higher values on graphene. Remarkably, in the first two days of culture, the axons were longer on graphene by 79% and 73%, respectively (Table 1). At day 3, reduced although significant differences were found, with the axons on graphene 17% longer than on glass. After day 3, the axons crossing the microchannels started to form complex networks that impeded to further quantify their length with accuracy.

| Substrate | DIV | Axonal length (µm) | s.e.m. | Percentage increase on graphene with respect to the control |
|---|---|---|---|---|
| Graphene | DIV1 | 114.6 | 8.0 | 79.3 % |
| Glass | | 63.9 | 5.7 | |
| Graphene | DIV2 | 305.2 | 19.2 | 73.6 % |
| Glass | | 175.8 | 10.7 | |
| Graphene | DIV3 | 419.6 | 22.0 | 16.6 % |
| Glass | | 359.7 | 14.9 | |

*Table 1. Axonal length on microfluidic chamber at different DIV. Average ± standard error of the mean (s.e.m.) of length values plotted in Figure 1b, and percentage of length increase on graphene with respect to glass at different DIVs.*

It should be noted that the same axonal length analysis at an early developmental stage was also performed on standard and graphene coated coverslips, yielding results that confirm preferential neurites outgrowth on graphene (Fig. S1a,b). Furthermore, this analysis revealed an organized pattern of DRGs grown on graphene already at DIV3, whereby somas are clumped together and dense axonal bundles depart radially from them (Fig. S1c,d). Knowing the influence of the substrate nanotopography on cell spreading [21,2>], we studied graphene surface morphology via AFM (Fig. S2). The flatness of the surface, with a root mean square roughness of about 1.8 nm, made us exclude the presence of features that can lead to discontinuous cell–substrate adhesion [21]. Moreover, as graphene crumpling was never observed in our analyses, we can exclude its effect on substrate wettability and soma clumping [21,23].

Obtained results support the trend previously reported for embryonic DRG neurons on epitaxial graphene [4]. Furthermore, the time-dependent axonal increase well agrees with what reported for non-compartmentalized hippocampal neurons on CVD graphene, where an initial ~30% increased elongation was reported for DIV2, followed by a stabilization at ~13% in the following days [1]. It is thus established that graphene strongly impacts the early development of axon elongation in peripheral sensory neurons. The optical transparency of graphene makes it an ideal substrate to probe possible changes in uptake and trafficking of trophic factors, which are at the basis of axon elongation and neural survival.

**Graphene alters retrograde transport of Nerve Growth Factor signaling endosomes**

Survival of sympathetic and sensory neurons is known to rely on the retrograde axonal transport of NGF signaling endosomes back to the cell body [17,18,2>], while numerous studies indicate that axon elongation is prompted by a local effect of NGF which does not directly involve mechanisms in the cell body [25]. Hence, we examined whether the observed graphene-induced axon outgrowth correlates with an altered NGF axonal transport. We used fluorescence microscopy to perform axonal transport studies in living, compartmentalized DRG neurons using a fluorolabeled NGF variant (fluoNGF) covalently coupled to an Alexa488 organic dye [26,27].

A schematic of the experiment is shown in Figure 2a.

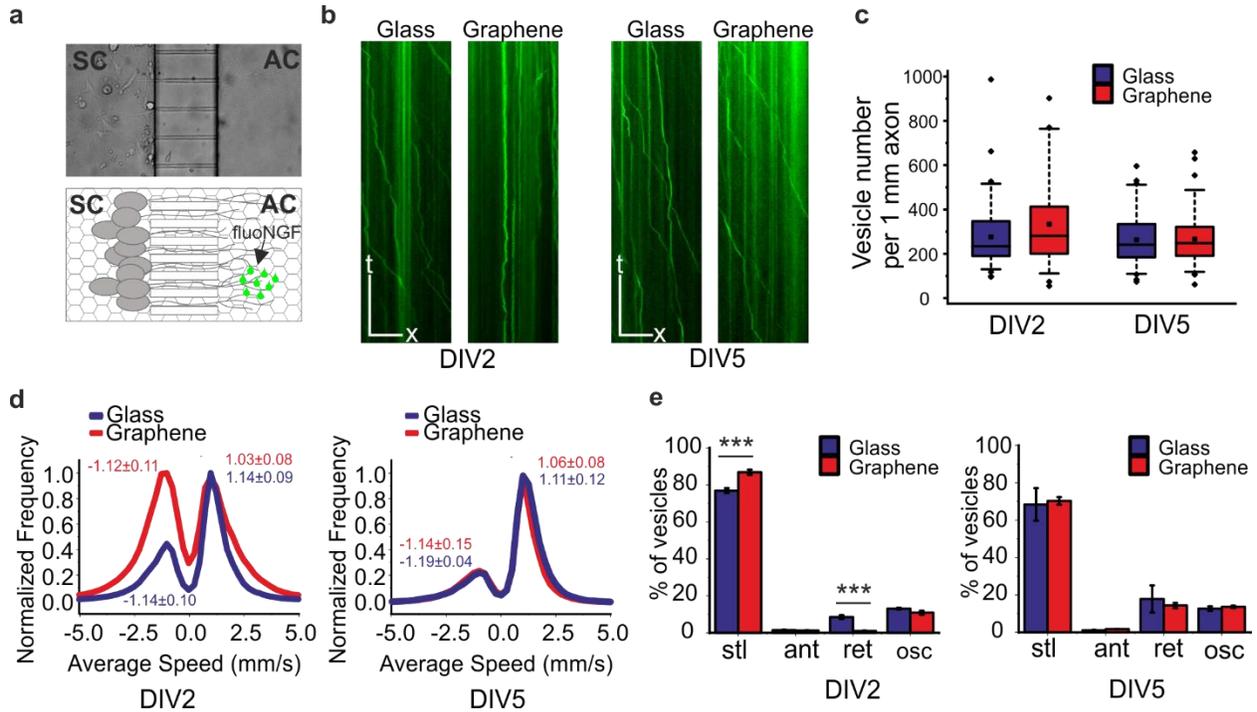

*Figure 2. Altered axonal transport of fluoNGF in DRG neurons on graphene (a) Schematic representation of compartmentalized microfluidic cell culture. Neurons are seeded in the soma compartment (SC) and extend their axon in the 150 μm microchannels reaching the axonal side (AC), where fluoNGF is administered (green droplets). (b) Representative kymographs of fluoNGF vesicles along a single axon at DIV2 and DIV5. x-scale bar, 5 μm; y-scale bar, 10 s. (c) Box plots for the number of vesicles per 1mm of axons after the administration of fluoNGF at the axon compartment of DRG neurons cultured on graphene and glass at DIV 2 and DIV 5. Data are not significantly different according to a Mann-Whitney test. In box plots: box between 25th and 75th percentile; horizontal line: median; whiskers: 5th and 95th percentiles; square: mean; circles:*

*outliers. (d) Speed distributions for moving parts of vesicle trajectories on graphene (red) and glass (blue) at DIV2 and DIV5. The mean ± standard deviation (SD) of each peak position calculated from the three independent replica is reported in the graph in the same color. Distributions with maxima normalized at 1. (e) Percentage of vesicles in different categories at DIV2 and DIV5 (mean ± s.e.m. for independent cultures). The four categories are: stalled (stl), retrograde (ret), anterograde (ant) and oscillating (osc). The percentage of stalled and retrograde vesicles on graphene at DIV2 were significantly different from the control (\*\*\*P<0.0001, One-way Anova, with Bonferroni's Multiple Comparison Test). The distribution of the populations at DIV5 did not significantly differ, according to One-way Anova. The number of acquired time-lapse images used to calculate the vesicle population was 81 for glass and 56 for graphene at DIV2 and 75 for glass and 72 for graphene at DIV5. For all panels, the number of vesicles at DIV2 in three independent cultures was 3604 for graphene and 5188 for glass. The number of vesicles at DIV5 in two independent cultures was 4905 for graphene and 5399 for glass.*

To assess the effect of graphene on the transported vesicles, NGF vesicular trafficking was examined within the microchannels, using compartmentalized neurons on glass as control. Visual inspection of the detected vesicles suggested that graphene induces a strong reduction of retrogradely transported NGF vesicles at DIV2 (Supplementary videos 1 and 2), while this is not the case at DIV5 (Figure 2b). To clarify if such change is due to an altered NGF uptake on graphene, we quantified the number of vesicles per mm of axons after NGF administration (Figure 2c) in control and graphene, and found no significant difference between the two groups, both at DIV2 and DIV5. This is in agreement with the reported unmodified neuron survival rate on graphene compared to the control [4]. We next ruled out that the observed effect was due to a slowing down of vesicles, and accordingly quantified the velocities during retrograde and anterograde motion of the vesicles on graphene and glass (represented by positive and negative velocities respectively, Figure 2d). We found that velocities of anterograde and retrograde transport were not altered by graphene both at DIV2 and DIV5. Also, we observed, similarly to what reported in [26,2>], that on glass vesicles underwent preferentially a retrograde movement, resulting in an asymmetric distribution. Surprisingly, on graphene we observed a symmetric distribution at DIV2, indicating comparable retrograde and anterograde movements for NGF vesicles. On the contrary, at DIV5 vesicles on graphene and glass exhibited similar distributions of speeds with predominant retrograde transport. The obtained data prompted us to analyze the entire population of NGF vesicles and classify them based on their movement. We distinguished four different categories: (1) stalled (stl), when the vesicles did not move; (2) retrograde (ret) when the vesicles moved from the axon tip to the cell soma; (3) anterograde (ant) when the vesicles moved from the cell soma to the axon tip; and (4) oscillating (osc) when the movement of the vesicle switched between retrograde and anterograde within a limited space. We found that in graphene and control the majority of NGF vesicles were stalled both at DIV2 and DIV5 (Figure 2e). However, on graphene at DIV2 the number of retrogradely transported vesicles was 8.5-fold reduced with respect to the control, in favor of a 13% increase of the stalled vesicles (Table S3). When comparing the oscillating and anterograde populations we found no significant difference between graphene and control. Hence, in the first developmental stage the graphene effect on NGF transport is to reduce retrogradely moving vesicles in favor of a locally stalled population. Importantly, this effect almost disappears in more mature DRG cultures, where we found that the percentages of both stalled and moving vesicles were superimposable to the control (Figure 2e right). This time-dependent fashion matches the predominant effect of graphene in axon elongation at early developmental phases (Figures 1 and S1), suggesting a key role for this trafficking alteration in the process. In order to elucidate the cause/effect links between graphene, NGF vesicle stall and axon elongation, we next investigated possible electrophysiological and structural changes prompted by the material-neuron interface.

**Altered neuron excitability and electrostatic interaction with graphene**

Electric activity has been recently associated to increased retrograde flux of signaling endosomes in hippocampal neurons [2] and to inhibited axon outgrowth in adult sensory neurons [30]. Hence, we carried out patch-clamp electrophysiology measurements to study the effect of graphene on DRG membrane resting potential and neuron excitability. To allow a direct comparison with the axonal elongation and transport studies, experiments were performed at similar time points. We observed a significantly hyperpolarized resting membrane potential for graphene-cultured DRG neurons with respect to glass-cultured controls regardless of the DIV investigated. Cellular excitability was further investigated by quantifying the number of spikes evoked by current injection steps. We observed that, at DIV2, the mean number of spikes in graphene was significantly lower than in the control (Figure 3b). However, this difference disappeared at DIV4, with control and graphene neurons firing a similar number of spikes (Figure 3c).

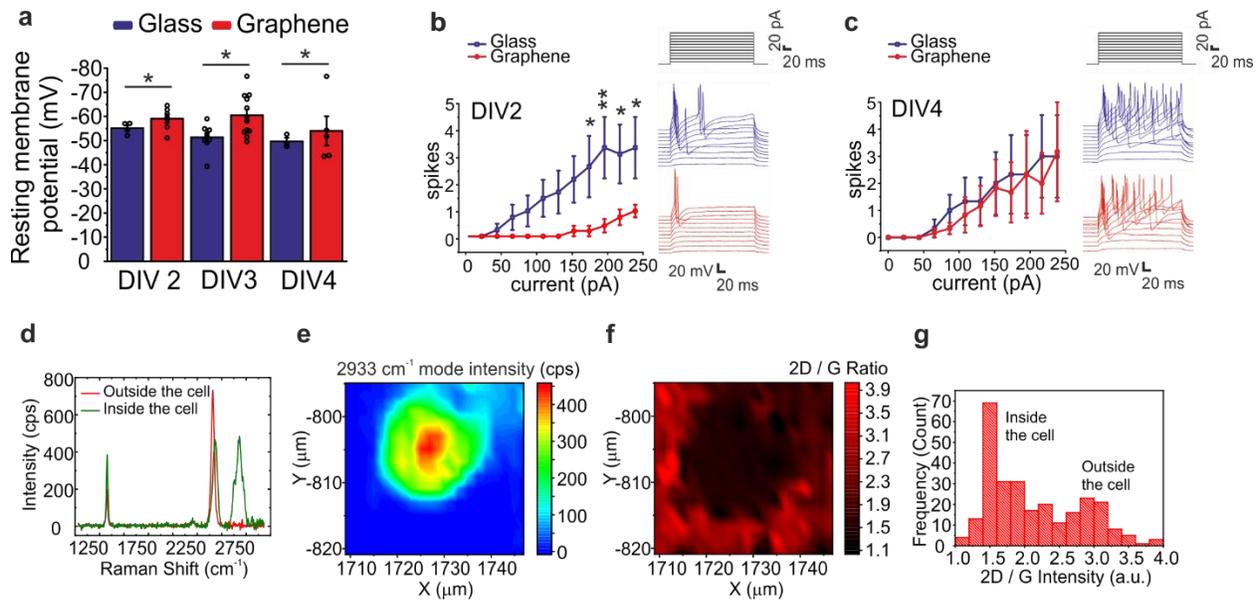

*Figure 3. Altered neuron excitability and graphene electronic properties*. (a) Graphene-cultured DRGs show hyperpolarized resting membrane potential in comparison to glass-cultured controls (*P_substrate <0.05, Two-way Anova). The number of measured neurons, collected in three independent cultures, was: DIV2, glass n=4, graphene n=7; DIV3, glass n=9, graphene n=12; DIV4, glass n=3, graphene n=5. (b,c) Number of spikes in response to current injection and representative traces on graphene and glass-cultured DRG at different DIVs; voltage traces at the different injected currents shown in the top graphs are shifted vertically. (b) On DIV2, graphene-cultured DRGs show reduced spike number in response to current injection with respect to glass-cultured controls (P_substrate X current <0.001, Two-way Anova, with Holm-Sidak post-hoc test, *p<0.05, **p=0.005). (c) On DIV4, no difference in spiking was observed between graphene- and glass-cultured DRGs according to Two-way Anova Test at 0.05 significativity. The number of measured neurons, collected in three independent cultures, was: DIV2, glass n=5, graphene n=6; DIV4, glass n=3, graphene n=6. (d) Representative Raman spectra obtained on bare graphene (red line) and on neuron/graphene interface at DIV2 (green line). (e) Raman map of the 2933 cm$^{-1}$ mode intensity, identifying the cell position. (f) 2D/G intensity ratio map of the same area

*reported in (d), revealing a shift of the ratio in the area underneath the cell. (g) Histogram of 2D/G intensity ratio showing the bimodal distribution of the intensity outside the neuron and the intensity underneath the neuron. The histogram reported in Figure 3g shows a bimodal distribution, where the distribution of 2D/G intensity ratio of bare graphene is peaked at 2.8 while the ratio in case of the graphene underneath the neuron decreases down to 1.5.*

These results demonstrate that the early days of contact with graphene are characterized by a change in neuron excitability, followed by a phase of adaptation. Hence, we describe a correlation between neuron excitability, axon elongation and signaling endosomes transport. The synchrony between restored excitability and recovery of normal NGF retrograde transport dynamics with prolonged culturing is a strong clue of such correlation. The recovery of normal signaling endosome transport shall, in turn, contribute to maintain neuronal survival [3]>.

In order to understand the reasons behind the observed reduced cellular excitability in the cell/electrolyte/graphene system, we carried out further electrical and spectroscopic analysis. Electrical measurements with a van der Pauw geometry of the graphene sample (i.e., coated and after medium immersion, with no neurons on top) indicate that holes are the majority carriers (Table S4). Peripheral neurons are known to be negatively charged, possessing more negative resting membrane potential with respect to central neurons [3]>. We thus investigated by Raman spectroscopy whether the local charge in graphene could be affected by a significant electrical interaction with the cell. Figure 3d-g shows the results obtained by carrying out Raman mapping of graphene with cultured DRG neurons at DIV 2. The Raman spectrum measured for the neuron/graphene system (green line) presents, in addition to the well-known G and 2D Raman modes [33], a complex band peaked at 2933 cm$^{-1}$, with one shoulder at 2885 cm$^{-1}$ (panel d). These Raman modes can be attributed to the cell and in particular to the $CH_3$ stretching (proteins) and to the $CH_2$ antisymmetric stretching (lipids), respectively [34]. Indeed, Figure 3e shows the Raman map of the 2933 cm$^{-1}$ mode intensity, identifying the cell position. The same areas were mapped to extract the 2D/G intensity ratio map (Figure 3f). The intensity ratio was found to be decreased precisely underneath the neuron (Figure 3g), demonstrating an increase of the local hole doping of graphene (from $2\times10^{12}$ cm$^{-2}$ to $6\times10^{12}$ cm$^{-2}$), clearly ascribable to the cell negative membrane potential. This value is in agreement with the one reported for a graphene-astrocyte system [35].

It is known that carbon-based π-electron-rich surfaces are ideal adsorption sites for potassium ($K^+$) ions [36,3>]. In particular, it was shown that the electronic properties of graphene can tune the efficiency of graphene in trapping $K^+$ ions, and that such trapping ability is likely to be significant in p-doped graphene (i.e., our case) [11]. This will reduce the extracellular concentration of potassium ($K^+$) ions, thus affecting transmembrane ionic currents. Thus, we speculate that a modified (i.e., reduced) extracellular $K^+$ concentration due to preferential adsorption of $K^+$ ions may increase $K^+$ currents and dampen neuronal excitability of graphene-cultured DRGs. At this time, the cause/effect link between altered excitability and modified NGF transport remains elusive, and further studies shall be performed to elucidate it. Since NGF transport could be also influenced by the axonal structure on graphene, in the next section we examine in the ultrastructure of axons grown on graphene at early developmental stage.

**Reduced microtubule distance and elongated axonal topology on graphene**

NGF retrograde axonal transport is mediated by the dynein-microtubule (MT) transport system [3>] and can thus be influenced by structural changes in the axon. This is maintained by a complex system of proteins including MTs and actin filaments that ensures shape, capacity to direct movements and extension of the growth cone [39].

Here, we first took the membrane-associated periodic skeleton (MPS) in consideration. This is a periodic ring-like structure around the axon circumference formed by actin, spectrin and associated proteins, found in a broad range of neuronal cell types including DRG neurons [40,4>]. We examined the MPS with stochastic optical reconstruction microscopy (STORM), for studying possible variations induced by graphene. DRG neurons were fixed at DIV2 and immunostained for βII spectrin for STORM imaging (Figure 4).

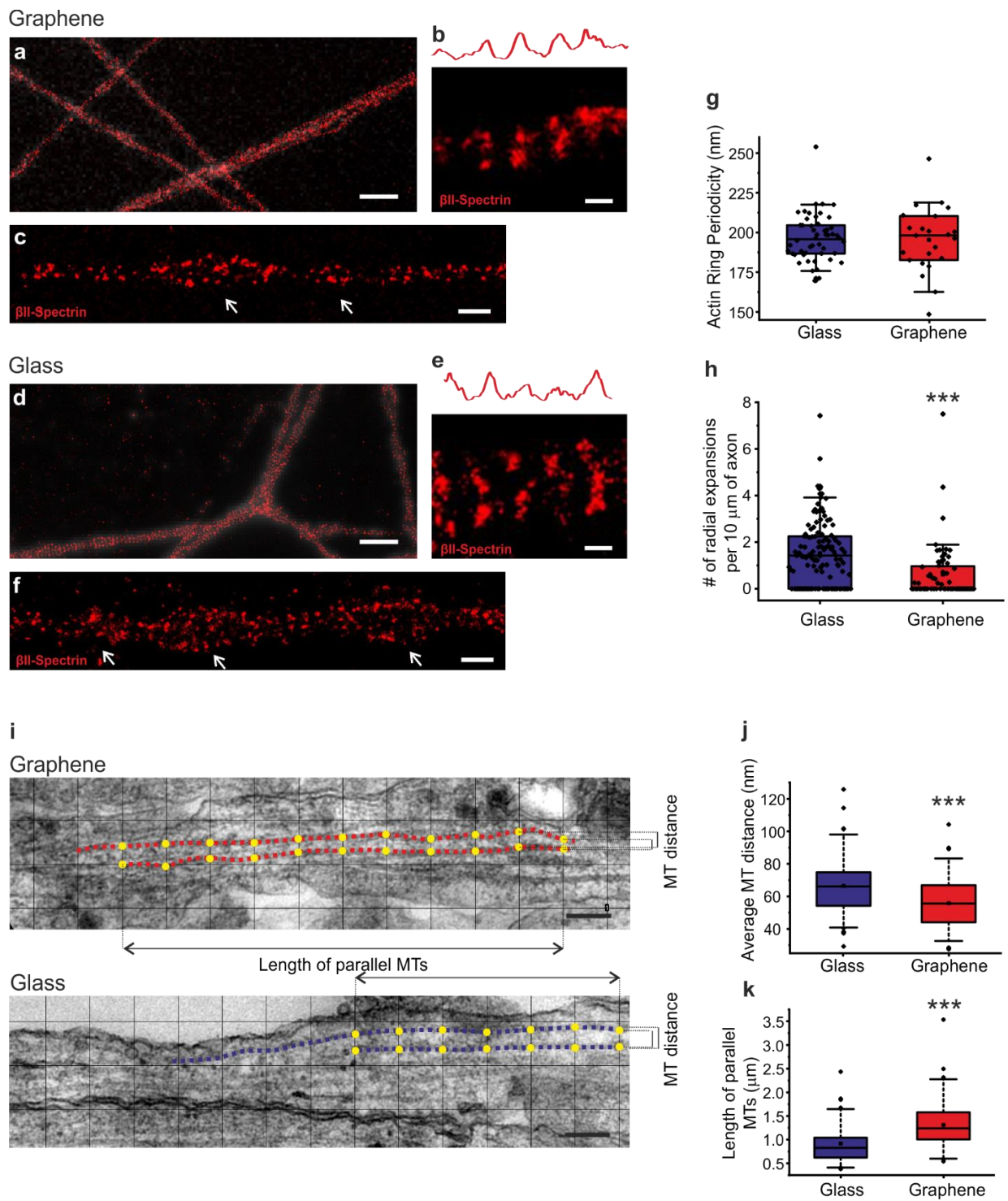

*Figure 4. Modifications of axonal topology and cytoskeleton of DRGs on graphene (a,d) Representative STORM images of βII spectrin (red) in DRG neuron cultured on glass and on graphene and fixed at DIV2, superimposed to conventional fluorescence images (grey). Scale bar:*

*20 μm. (b,e) Magnification of the STORM images with intensity profiles across the periodic spectrin structure reported with a red line. Scale bar: 200 nm. (c,f) Magnification of the STORM images with white arrows indicating MPS radial expansions (Scale bar: 500 nm) (g) Box plots showing the periodicity of spectrin ring-like structure, which did not significantly differ according to Mann Whitney Test. The number of analyzed regions is 21 for graphene and 48 for glass, each from two independent cultures. (h) Box plots of the number of radial expansions per 10μm of axon, showing a significantly higher number of radial expansions on glass than on graphene (\*\*\*P<0.0001, Mann Whitney Test). The number of analyzed axons is 64 for graphene and 123 for glass, each from two independent cultures. For the box plots in g) and h): box between 25th and 75th percentile; horizontal line: median; whiskers: $5^{th}$ and $95^{th}$ percentiles; square: mean; circles: measured values. (i) Representative TEM images for quantifying axon diameter and number, geometry, and spacing of MTs of DRG cultured on graphene and glass. Scale bar: 200 nm. A couple of adjacent MTs is highlighted with dotted red lines (graphene) or with dotted blue lines (glass). Grid-pattern lines are superimposed in black and the intersection points of the grid lines with the MT, used to quantify the MT distance, are highlighted in yellow. The length of parallel MTs and the MT distance are also reported. (j) Box plots showing the average distance between a couple of microtubules within axons on glass and graphene (\*\*\*P<0.0001, Mann Whitney Test). (k) Box plots showing the lengths of parallel sections of microtubule couples inside the axon on glass and graphene. For the box plots in j) and k): box between 25th and 75th percentile; horizontal line: median; whiskers: $5^{th}$ and $95^{th}$ percentiles; square: mean; circles: outliers; number of analyzed couples: 79 for graphene and 97 for glass, each from two independent cultures.*

We found that graphene-cultured neurons maintain the characteristic periodicity of ~190 nm of the MPS (graphene, 195.6 ± 20.8 nm vs. glass, 196.7 ± 14.9 nm, Figure 4g). However, the super-resolution approach also allowed us to distinguish the presence of regions in which the MPS expands radially (white arrows in Figure 4c,f). These are compatible either with zones of minor compaction within axonal bundles [4], or with increased diameter of single axons. The number of MPS radial expansions was quantified over an axonal length of 10 μm and we found that on graphene they were more than halved with respect to the control (Figure 4h). These data confirm the observation that on graphene axonal bundles are tightly organized already at the early developmental stage (Figure S1c,d), but also suggest that axons therein may have a stretched structure, which would well correlate with their increased length (Figure 1).

We next exploited transmission electron microscopy (TEM) to analyze whether and how the MT organization is affected by this topological rearrangement. In detail, we quantified the spacing between adjacent couples of MTs in longitudinal sections of the DRG culture on graphene in comparison with the control (Figure 4i). We found that on graphene the intratubular distance was significantly reduced by 19%, when compared to the control (Figure 4j). In order to exclude an effect of the axon caliber on this measure [4], we quantified the mean axon diameter and the number of MTs per axon, which were found to be comparable in the two cases (Figure S5). Accordingly, a parallel immunofluorescence quantification of tubulin showed that the mean fluorescence intensity of Alexa488-labelled βIII-tubulin is similar for DRG cultures grown on graphene and control (Figure S6). Moreover, we measured the length for which the pairs of MTs arranged into adjacent arrays remains parallel along the same axon, finding that this parameter exhibits a remarkable 42% increase in graphene with respect to the control (Figure 4k), indicating that MTs are closer and straighter in axons on graphene than on the control substrate. This can be easily explained by the electrostatic interaction experienced by axons on graphene (Fig. 3), which may increase their adhesion to the substrate, finally changing their three-dimensional architecture and forcing them to remain straighter than on the control substrate. Overall, our results suggest that the interaction between graphene and neurons does not dramatically alter the axonal ultrastructural organization, maintaining a similar total number and integrity of MTs and MPS, but rather confers to axons an elongated morphology. This effect may also provide another cause for the observed stall of NGF signaling endosomes (Figure 2). Indeed, transient radial expansions of axonal diameter were recently described to facilitate the processivity of the fast-moving retrograde carriers [44]. Accordingly, the inhibition of these expansions prompted by graphene (Figure 4h), along with the reduced MT interdistance (Figure 4j), might create a more crowded space that inhibits cargo mobility and reduces the retrograde transport of NGF-loaded endosomes, ultimately favoring their local action on axon outgrowth.

**CONCLUSIONS**

Graphene has attracted the interest of the neuroscience community worldwide for its biocompatibility, electrical properties and regenerative potential [45–4>]. However, the real possibility of using this material for neuroregeneration applications will depend on our understanding of how graphene influences the processes of neurite outgrowth, elongation and regeneration following injury. This can be achieved by exploiting graphene optical transparency to observe the processes of interest, as we propose in this work. Optical microscopy of DRG neurons shows an axonal length greater on graphene than in the control. Single-molecule fluorescence microscopy in living DRG neurons reveals that, although on graphene neurons are completely viable for weeks [4,11], deep nanoscale changes characterize the early cell-material interaction. In detail, we observe an 8.5-fold reduction of retrogradely transported vesicles of NGF, the signaling molecule known to mediate sensory axon development (Figure 2e). This data is in line with previous observations that local NGF actions account for axon elongation [25]. Indeed, NGF-induced axonal sprouting in sensory neurons has been widely investigated, and NGF treatment probed in models of spinal cord injury [49]. Several aspects still need to be addressed, e.g. whether graphene locally stalls only NGF vesicles, or also other fast-retrograde carriers, if NGF internalized at the soma is also influenced by graphene, and how the spatiotemporal signaling events triggered by NGF receptors [50] change on graphene. Nevertheless, our experiments provide a clear clue that the physical control of the intracellular NGF pool may be useful for regeneration approaches.

To investigate possible causes for the observed NGF trafficking alteration, we studied graphene influence on both electrical and structural properties of axons. In the former case, we observe mutual effects: on one side neurons change graphene charge concentration, inducing a doping clearly measurable via Raman spectroscopy (Figure 3f); on the other side graphene hyperpolarizes the resting membrane potential, and reduces cell excitability (Figure 3a-c). It is noteworthy that this condition correlates with a maximal effect of graphene on axonal length (up to >70% increase with respect to a control substrate, Tables 1 and S1), and was previously linked to increased axon regeneration in sensory neurons [5]>. On the other hand, graphene also alters axon morphology, inducing longer and straighter axonal bundles (Figures 1, S1, 4h) and closer and straighter MTs within each axon (Figure 4j).

The sum of these effects is progressively diminished with time, allowing to reach already after 4-5 DIV an excitability and axonal transport comparable to the control (Figures 2e, 3b), and stabilizing to a 17%-increased axonal outgrowth (Figure 1b and Table 1). Thus, the molecular players stalling NGF signaling endosomes are selectively switched on upon the early contact with the material. More studies are needed to understand if the electrophysiological and structural graphene effects can impact independently the transport of NGF vesicles, or if they are rather synergistic. This will depend on the ability to experimentally dissect membrane hyperpolarization, axonal stretching, and retrograde transport inhibition, while not precluding the correct NGF uptake to preserve neuron survival. Obtained results may surely increase our mechanistic detail of axonal growth processes and help to design optimal biocompatible guides to enhance them.

**EXPERIMENTAL METHODS**

**CVD graphene synthesis and transfer**

The graphene was synthesized using an Aixtron BM Pro cold-wall reactor on electropolished Cu foil (purity 99.99%, Alfa-Aesar) as previously describe in ref [52,53]. The copper foil was annealed in argon for 10 min to reduce the terraces to uniform the growth of polycrystalline graphene. The growth was performed introducing methane and hydrogen for 30 min at a temperature of 1020-1040 °C and a pressure of 25 mbar. The gas flow rates were typically 2 sccm of methane, 20 sccm of hydrogen and 980 sccm of argon. Average domain size of graphene was 20-30 μm.

The transfer of single layer graphene (SLG) on glass coverslips was performed using the standard wet etching technique [52,54]. A layer of poly-methyl-methacrylate (PMMA, 679.04) was spin coated on the SLG/Cu foil at 4000 rpm for 1 min and heated for 1 min at 90 °C. The graphene grown on the back-side of the Cu foil was removed by reactive ion etching (RIE), using 80 sccm $O_2$, 20sccm Ar, 1 min. The Cu was etched overnight using a 0.1 M solution of ammonium persulfate (APS). The membrane was repetitively rinsed in deionized (DI) water to remove any residuals of the Cu etchant. PMMA/SLG membrane was transferred on a glass coverslip, previously cleaned with acetone and isopropanol and treated with oxygen plasma for 5 min at 100 W. When completely dried, the sample was kept in acetone for at least 2 h to remove the PMMA, rinsed with isopropanol and annealed at 170 °C for 3 h on a hotplate.

**Surface functionalization**

PDL with laminin coating (30µg/ml PDL (P1149 Sigma) and 5µg/ml laminin (Life Technologies)) was selected to carry on the DRG culture experiments. An incubation time of 4 h was used for coating the 10 mm round glass coverslip [4]. When microfluidic chambers were used, an overnight incubation was chosen as suggested by the company.

**Raman measurements**

The quality of graphene after its transfer on glass was assessed by Raman spectroscopy. A micro-Raman spectroscope (inVia Raman, Renishaw), equipped with a motorized sample stage and a 532 nm laser with a spot size of around 1 µm in diameter, was used to map the characteristic graphene Raman peaks. The laser power was kept at 1 mW with an acquisition time of 2 s. The Raman spectra confirmed the single layer nature of the graphene samples, with prominent G and 2D peaks, narrow 2D peak (FWHM = 29-34 cm$^{-1}$) and absence of the D-peak [55,56].

**Harvesting and dissociation of dorsal root ganglia**

All animal procedures were approved by the Italian Ministry of Health (notification n° 917) and were fully compliant with Italian (Ministry of Health guidelines, Legislative Decree n° 26/2014) and European Union (Directive n° 2010/63/UE) laws on animal research. The experiments were carried out in strict accordance with the approved guidelines. In addition, the principles of the Basel Declaration, including the "3R" concept, have been considered throughout the whole project. The primary sensory neurons used in this study were obtained from the dorsal root ganglia of wild-type B6129 P3-P4 neonatal mice, by adopting the protocol described in Testa *et al.* [57]. The dissection was performed under a dissection hood. Briefly, the mice were killed via cervical dislocation and the dorsal side of the spinal cord was incised and lifted until reaching the caudal part. The spinal marrow was removed to locate the DRGs. The DRGs were carefully removed using a stereo microscope and collected in a petri filled with a solution of PBS and 50 units/ml of Pen-Strep on ice. 20-30 DRGs were collected per animal from all spinal levels.

The DRGs were enzymatically digested by incubating them for 40 min at 37 °C in a PBS solution containing 0.03 % Collagenase from Clostridium histolyticum (C7657 Sigma), 0.3% Dispase II protease (D4693 Sigma) and 0.18% glucose. After digestion, ganglia were transferred to a 50 mL conical centrifuge tube with 5 mL of PBS containing 0.01% Deoxyribonuclease I from bovine pancreas (D5025 Sigma) and 0.05% Trypsin inhibitor from Glycine max (T9003 Sigma). Ganglia were dissociated by mechanical agitation through a fire-polished glass Pasteur pipette until the suspension was homogeneous. The solution was then centrifuged for 10 min at 1000 rpm and resuspended in Primary Neuron Basal medium (PNBM, Lonza) supplemented with 1% L-glutamine (Lonza), 0.1% Gentamicin Sulfate/Amphotericin-B (Lonza), 2% NSF-1 (Lonza).

For the survival of DRG neurons, 100 ng/ml of NGF (Alomone Labs) were added to the media. Since 24 h after seeding, 2.5 μM AraC (Sigma) was added for inhibition of glia proliferation. Half of the medium was replaced every 3–4 days. Axonal length, transport and electrophysiology studies were performed on live neurons; immunofluorescence and Raman studies were performed on neurons fixed in 4% v/v paraformaldehyde (PFA) in PBS.

**Microfluidic cell culture platform**

For compartmented cultures, PDMS microfluidic chambers (RD450 and RD150, Xona Microfluidics) were used. Bare glass coverslips were treated overnight with 65% nitric acid (Sigma), washed several time with DI water and left in 96% ethanol. Prior to assembly the glass substrates with the chamber, both the coverslips and the devices were immersed in 96% ethanol for 30 min, rinsed several times with DI water and dried under a tissue culture hood to guarantee sterility. When graphene was used, a CVD-grown graphene sheet was transferred on the glass before the device assembling. The device was then assembled by simply placing it on top of the coverslip and pressing down with the back of a pair of tweezers to create a seal between the PDMS and the substrate. It was washed with ethanol and DI water. Due to the high hydrophobicity of the device and glass a micropipette was used to push the water into the main channel and remove any possible bubbles. Next, the coating solution (150-200 µl, 30 µg/ml PLD and 2 µg/ml laminin in DI water) was added to each compartment and incubated overnight at 37 °C. The sample was cleaned repeatedly with DI water and incubated with DI water for 1 hour. It was then washed for the last time with the cell culture medium before cell seeding.

**Neuron seeding**

Neurons seeded on the coverslips were plated as a drop of 80 µl on each coverslip and left to attach for 1 h, before 500 µl of culture medium were added.

When a more homogeneous cell distribution was preferred (e.g. low cell density for axonal elongation measurements), neurons were suspended in a larger volume of culture medium and 500 µl of culture medium were added to each well.

When neurons were plated in microfluidic devices, 10 µl of neurons were loaded in the soma compartment at a concentration of 2.5 to 4.5 million cells per ml. The device was incubated for 1 h to allow the cell to attach and then each well was filled with the media. The medium was changed every 1-2 days to prevent salt concentrations increasing due to medium evaporation. Moreover, after 24 hours from the seeding, 50 ng/ml of NGF were added to the soma compartment and 100 ng/ml to the axon compartment to generate a gradient of NGF and induce DRG axons to reach the axonal side by crossing the microgrooves [26]. A volume difference between the two compartments (~50 μl, higher volume on the soma side) was applied to fluidically isolate them [58].

**Axonal length quantification**

Axon lengths for neurons seeded in the microfluidic devices were measured daily in the microgrooves compartment and in the axon compartment. The length quantification was performed excluding the axonal segments in the soma compartment due to impossibility to distinguish single axons in the network complexity. We performed two independent cultures and quantified more than 50 axons per sample. Transmitted light imaging were acquired at 37 °C, 5% $CO_2$ in differential interference contrast (DIC) configuration using an inverted Leica AF6000 microscope with an oil immersion 63x objective. For axon length quantification of neurons seeded on glass coverslips, we performed two independent cultures with three and four biological replicate each. We analyzed at least 40 cells from more than 10 selected fields obtained with an inverted microscope equipped with a 20× objective (Leica DMI4000B microscope).

**Immunofluorescence staining**

Neurons were fixed at various DIV using 2% (w/v) Paraformaldehyde (PFA), 4% (w/v) sucrose in PBS for 20 min at room temperature (RT) and rinsed 3 times with PBS. Fixed neurons were then permeabilized in 0.1% (v/v) Triton-X-100 in BSA 2% (w/v) in PBS for 7 min. After 5 washes in PBS, samples were blocked in blocking buffer (BSA 5% in PBS) for 1 h and subsequently stained with mouse anti-NfH antibody (1:500, neurofilament heavy chain, Abcam ab7795) in incubation solution (BSA 2.5% in PBS) for 2 h. After PBS washing, the samples were stained for 1 h at RT with secondary antibodies anti-mouse-Alexa488 (1:100, Invitrogen, A21202) diluted in incubation solution. Coverslip were mounted on glass slide in Fluoroshield with DAPI mounting media (Sigma).

Alternatively, neurons were stained with primary mouse anti-beta III Tubulin antibody (1:200, Abcam, ab78078), with secondary antibodies anti-mouse-Alexa488 (1:100, Invitrogen, A21202) diluted in incubation solution for 1 h and then with Alexa-647 conjugated phalloidin (500 nM in PBS, Invitrogen A22287) overnight at 4 °C. Coverslip were mounted on glass slide in Vectashield mounting medium (Vector Laboratories).

For the fluorescent labelling for STORM imaging, we used the protocol reported in He *et al.* [41] and Han et al. [59]. Neurons were fixed at various DIV using 4% (w/v) PFA in PBS for 20 min at room temperature (RT) and rinsed 3 times with PBS. Fixed neurons were then permeabilized in 0.2% (v/v) Triton-X-100 in in PBS for 5 min. Permeabilized samples were incubated for 1 h in blocking buffer (BSA 3% w/v in PBS) and finally stained with primary antibody to label βII spectrin (1:300, BD Biosciences, 612563) in blocking buffer overnight at 4 °C. As secondary antibody we used anti-mouse-Alexa647 (1:100, Invitrogen, A31571). Coverslip were mounted on glass slide in Vectashield mounting medium (Vector Laboratories).

**FluoNGF production**

To label the neurotrophin with the fluorophore we applied a method firstly described in Yin *et al.* [60] and optimized as in Di Matteo *et al.* [27]. We used a modified neurotrophin, obtained by fusing an 11-residue peptide named ybbR with the C-terminus of the NGF sequence [27]. The NGF-ybbR was adopted for site-specific protein labeling by Sfp synthase enzyme.

For the fluorolabeling reaction, 45 µg of NGF-YBBR were diluted in PBS and spun at 7000 rpm for 1 min at 4 °C. The following reagents were then added: 73 µM CoA-Alexa488, 17µM Sfp Synthase, 36 mM $MgCl_2$ in PBS up to 135 µl final volume. The reaction was performed in the same tube were the NGF was stored. The solution was incubated at 37 °C and 350 rpm for 30 min. When the reaction was completed, the tube was placed in ice before the purification via HPLC.

We performed a cation exchange HPLC, using a Propac SCX-20 column (Dionex, Thermo Fisher Scientific) with 100 mM $HNa_2PO_4$ mobile phase with 0 M to 1 M NaCl gradient. The temperature was kept at 4 °C. The elution time of the reacted neurotrophin was shifted due to the negative charge of the fluorophore. The fluorescent NGF (fluoNGF) was separated from the non-reacted neurotrophins and from the free fluorophore by measuring the absorbance at 280 nm and 488 nm. The fluoNGF was stored at 4 °C for at most 15-20 days.

**Transport measurements**

To image transport of fluoNGF in live DRG neurons, the compartmentalized culture was supplied with 2nM of fluoNGF in the axon side and incubated for 1 h. The sample was then mounted on a homemade microscope stage and kept at 37 °C and 5% $CO_2$ during measurements. Time-lapse images were acquired using an inverted epifluorescence microscope (Leica AF6000) equipped with Leica TIRF-AM module, a 100× oil immersion objective (NA 1.47), and a Hamamatsu camera (ImagEM C9100-13, Hamamatsu).

NGF conjugated with Alexa 488 were excited with a 488 nm laser in epifluorescence configuration, with an exposure time of 100 ms. Up to 1000 frames for each experiment were acquired. Live cell fluorescence imaging was effectuated in the channel compartment, after 30 min from the neurotrophin administration, for about 45 min.

**Single particle tracking analysis**

Tracking of single vesicles containing NGF was performed using homemade scripts in MatLab (The MathWorks), starting as described in De Nadai *et al.* [26]. Briefly, we selected manually the channels present in each field of view, and used functions distributed by Raghuveer Parthasarathy [61] for sub-pixel localization and tracking (https://pages.uoregon.edu/raghu/particle_tracking.html). Trajectories longer than 5 frames were considered and divided in subtrajectories on the basis of the vesicle velocity calculated using a mobile window of 5 frames, considering retrograde (positive) and anterograde (negative) movement for velocities higher than 0.5 µm/s or lower than -0.5 µm/s, respectively. We calculated the velocity distribution of these mobile subtrajectories, weighting the velocity of each subtrajectory with its time length and considering its s.e.m. as in [26]. Differently from [26], in order to quantify also the stalled NGF population, we did not apply any threshold in the speed or in the distance covered by the vesicles. Rather, we used all the trajectories, and classified them in 4 populations based on the way they moved along the channel. For this classification, trajectories shorter than 30 frames were discarded. Vesicles were classified as "stalled" when: a1) being composed by less than 10% of moving subtrajectories and a2) displaying a total average anterograde or retrograde speed less than 0.03 µm/sec; or b) displaying a total displacement less than 1.14 µm. Vesicles showing an oscillatory transport with a succession of anterograde and retrograde motions and average total speed less than 0.1 µm/sec, or with total shift less than 2.28 µm, were classified as "oscillating". The other trajectories were considered "retrograde" or "anterograde" depending on the direction of the total shift, respectively from the axons to the soma or vice versa.

**Patch clamp recordings on DRG cultures**

Recordings were performed on DRG primary neuronal cultures, prepared as described above, by adapting the procedure described in Siano et al.[62]. Cells were continuously bathed using Tyrode's solution containing (in mM): NaCl 150, KCl 4, $MgCl_2$ 1, CaCl2 4, Glucose 10, HEPES 10, pH 7.4 with NaOH. Borosilicate glass pipettes were pulled with a P-97 puller (Sutter, CA) to a resistance of 5-6 MΩ when filled with an internal solution containing (in mM): K-Gluconate 145, MgCl2 2, HEPES 10, EGTA 0.1, Mg-ATP 2.5, Na-GTP 0.25, phosphocreatine 5, pH 7.35 with KOH. After achieving a gigaseal and, then, whole-cell configuration, at least 3 min were allowed for complete equilibration between cytosol and internal solution. After switching to I=0 pA configuration, the resting membrane potential was measured for at least 2 min. Subsequently, in current clamp configuration, current injection was adjusted to obtain an initial membrane potential of -60 mV before delivering 20-pA current steps. Access resistance and membrane capacitance were monitored, and recordings were accepted only if series resistance had varied less than 20% of the initial value. Data were acquired using a MultiClamp 700A amplifier, connected to a Digidata 1322A digitizer (Molecular Devices, CA). Data were analyzed using Clampfit 10.7 (Molecular Devices).

**STORM sample imaging and data analysis**

STORM microscopy was performed using a Nikon N-STORM system configured for total internal reflection fluorescence imaging. Alexa-647-labelled samples were excited by 647 nm laser line. The ultraviolet (405 nm) line was used to activate the photoswitchable dyes during frames acquisition.

Fluorescence was collected using a Nikon 100x, 1.4NA oil immersion objective and passed through a quad-band-pass dichroic filter (97335 Nikon). Time-lapses of 30000 frames were recorded on a 256 × 256 pixel field (pixel size 160 nm) of an EMCCD camera (ixon3, Andor). Data were analyzed with the STORM module of NIS Elements (Nikon), ImageJ and Origin.

We selected a ROI in each image containing a portion of the axon with more than 5 ring-like structures and used a Fourier transform of the intensity profiles of spectrin along the axon to calculate the fundamental frequencies corresponding to the spatial period. The intensity profiles showed in Figure 4(b,e) were extracted from the ROI using the "Plot Profile" tool in ImageJ.

**TEM**

DRG neurons (DIV3) were processed as described before [63]. Briefly we fixed cells on coverslips with 1.5% glutaraldehyde in 0.1 M sodium cacodylate buffer, pH 7.4, post-fixed in 1% $OsO_4$, 1% $K_3Fe(CN)_6$, 0.1 M sodium cacodylate buffer, stained with our homemade staining solution [64], dehydrated in a growing series of ethanol, and flat-embedded in Epoxy resin. Ultra-thin sections (80 nm) were cut with a LEICA UC7 and imaged with a Transmission electron microscope Zeiss LIBRA 120 Plus operating at 120 KeV equipped with an in-column omega filter; images were recorded at different magnification. We imaged more than 30 ultra-thin sections covering a thickness of around 4 μm. The images related to two independent experiments were morphologically and morphometrically analyzed using NIH ImageJ software.

For microtubule quantification, more than 200 axons were examined for graphene and control group, taken from more than 25 TEM images per sample. We observed MTs running adjacent and used them to measure the intratubular distance at regular intervals. We measured the MT spacing by taking a pair of adjacent microtubules and following them along the axon, monitoring their distance at regular points of 200 nm using a grid.

**Statistical Analysis**

We pooled together all values from samples of the same experiment. Data were analyzed by using Origin Software. ANOVA with Bonferroni multiple comparison test or Holm-Sidak post-hoc test for parametric data and by Mann-Whitney for non-parametric data were used for statistical significance with $*p < 0.05$, $**p < 0.01$ and $***p < 0.001$. Data are presented as mean ± s.e.m. unless otherwise specified.

## AUTHOR CONTRIBUTIONS

Coordinated the project: CC and LM. Designed the study: LM, CC and DC. Synthesized and characterized graphene substrates: NM, DC, CC. Established DRG cultures on graphene: DC, LM, GT, SC. Performed axonal transport measures and data analysis: DC, SL, LM. Performed Raman studies and data analysis: FF, DC, CC. Performed electrophysiology studies and data analysis: MM, DC, LM. Performed TEM experiments and data analysis: VC, DC. Performed STORM experiments and data analysis: DC, LA, LM, SL. Wrote the manuscript: DC, LM, CC, with contributions from all authors.

## ACKNOWLEDGMENT

We thank G. Signore and A. Moscardini for assistance in the NFG purification, R. Amodeo and L. Ceccarelli for Sfp synthase enzyme production and MT. Ciotti for the DRG harvesting and dissociation protocol. We thank A. Cattaneo, F. Beltram, V. Raffa and C. Leterrier for useful discussions.

**Supplementary results on graphene impact on axonal elongation**

The effect of graphene on axon outgrowth in primary peripheral neurons was investigated also with a standard approach, using neurons seeded on graphene transferred on glass coverslips (Figure S1). Neurons plated on glass coverslips were used as controls.

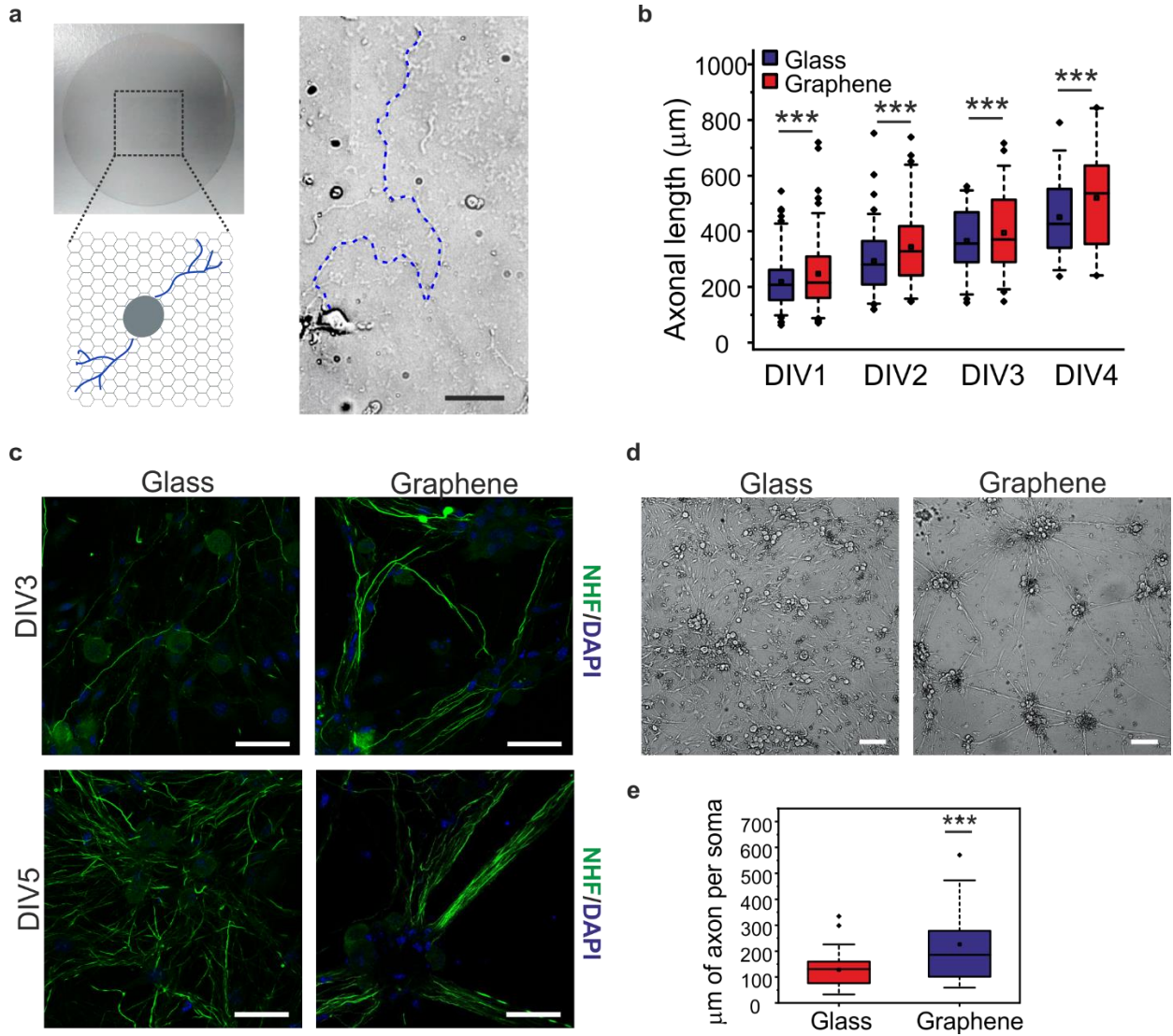

*Figure S1. Morphology and axonal elongation in DRG neurons grown on graphene (a) Schematic representation of DRG axonal length quantified on graphene transferred on a glass*

*coverslip. On the left, optical image of CVD grown graphene transferred on glass. On the right, the bright field image shows a neuron with axon marked with a blue dashed line. Scale bar: 50 µm. (b) Quantification of the length of measurable axons at different days in vitro (DIV) on glass and graphene. A significant difference is found between graphene and control according to Two-way Anova Test. (P_substrate <0.001, Two-way Anova, with Holm-Sidak post-hoc test, \*\*\*p<0.001 ). The number of measured axons was: DIV1, glass n=109, graphene n=106; DIV2, glass n=84, graphene n=70; DIV3, glass n=44, graphene n=51; DIV4, glass n=23, graphene n=33. (c) Fluorescence microscopy images of DRG neurons at DIV3 and DIV 5 stained for neurofilament heavy chain (NHF) to identify axons and DAPI to identify nuclei. Scale bar: 50 µm (d) Representative bright field images showing DRG neuron morphology on glass and graphene after DIV3. (e) Box plots of the axonal density in DRG neurons at DIV3 stained for NHF and DAPI. The number of acquired fields containing axons was 33 for glass and 31 for graphene, showing a significantly higher axonal length on graphene than glass (\*\*p<0.01), Mann Whitney Test). In box plots: box between 25th and 75th percentile; horizontal line: median; whiskers: $5^{th}$ and $95^{th}$ percentiles; square: mean; circles: outliers.*

Neurons were seeded at a very low density to distinguish single axons, as reported in Figure S1a. The length was quantified by manually tracing the axon from the soma to the axonal tip. We were able to quantify the neurite length up to 4 days of culture, as later on the neuronal network was too dense to distinguish isolated axons. Indeed, especially in more dense cultures, already after 3 days of culture, cells on graphene displayed a more complex distribution, a network of small cell-body aggregates connected by axons, differently from the control surface where they distributed homogeneously (Figure S1c,d). This is even more evident after axon immunostaining using anti-neurofilament NHF antibody (Figure S1c). We quantified also the axonal density, calculated as the ratio of the total axonal length in the field and the number of soma (Figure S1e). As expected graphene showed a significantly higher axonal density (Mann Mann Whitney, **P <0.01).

Figure S1b reports the axon length distribution measured on a coverslip at four time points. We found gradually augmenting neurite length both on graphene and glass, with a significant increase on graphene. The maximal elongation was observed at 2 days in vitro (DIV) where the axons were up to 17% longer than the control. For the details, see Table S1. In the compartmentalized approach, as reported in Figure 1b in the main text, the measured axon elongation was significant slower both at DIV1 and DIV2 with respect to the approach reported in this figure, both for graphene and control (Two-way Anova Test. (***P <0.001)) (Table S2). Differently from microfluidic chamber, on coverslip the percentage increase of axonal length on graphene with respect to control does not decrease with DIV. This could be due mainly to two reasons: i) axon growth on the device is slower (see absolute average length values reported in Tables 1 and S1 in the main text), because NGF stimulation is performed in a controlled way exclusively on the most distal part of the axon; ii) axonal length in the first days can be measured directly in the microchannel, where it has a linear growth trend. On the contrary, the length quantification on the coverslips is likely to be less precise, due to the faster axon growth triggered by the ubiquitous NGF stimulation and the presence of neural networks that already in the first two days make it harder to distinguish single axons, especially the longer ones, characteristic of neural growth on graphene. This is further confirmed by the axonal density reported in Fig.S1(e), in which thank to the immunostaining the quantification of all axonal branches in the coverslip allowed to obtain values comparable with the ones obtained on microfluidic chamber.

*Table S1. Axonal length on coverslip at different DIV.* Average length ± s.e.m. values represented in the box plot in Figure S1b, with the percentage of length increase on graphene with respect to the control at different DIV.

| Substrate | DIV | Axonal length (µm) | s.e.m. | Percentage increase on graphene with respect to the control |
|---|---|---|---|---|
| Graphene | DIV1 | 246.6 | 12.2 | 13.4 % |
| Glass |  | 217.5 | 9.1 |  |
| Graphene | DIV2 | 343.0 | 15.9 | 17.3 % |
| Glass |  | 292.4 | 12.7 |  |
| Graphene | DIV3 | 394.4 | 19.9 | 8.1 % |
| Glass |  | 364.7 | 17.4 |  |
| Graphene | DIV4 | 519.3 | 32.7 | 15.4 % |
| Glass |  | 450.1 | 30.3 |  |

*Table S2. Axonal length of DRG cultured on coverslip vs microfluidic chamber at different DIV.* Comparison between the length values reported in Figure 1b and Figure S1b, with the percentage of length increase on standard approach involving the coverslips with respect to the microfluidic chamber.

| DIV | Substrate | Approach | Axonal length (µm) | s.e.m. | Percentage increase on coverslips with respect to microfluidic chamber |
|---|---|---|---|---|---|
| DIV1 | Glass | Chamber | 63.9 | 8.0 | 240.4 % |
|  |  | Coverslip | 217.5 | 9.1 |  |
|  | Graphene | Chamber | 114.6 | 8.0 | 115.2 % |
|  |  | Coverslip | 246.6 | 12.2 |  |
| DIV2 | Glass | Chamber | 175.8 | 10.7 | 66.3 % |
|  |  | Coverslip | 292.4 | 12.7 |  |
|  | Graphene | Chamber | 305.2 | 19.2 | 12.4 % |
|  |  | Coverslip | 343.0 | 15.9 |  |
| DIV3 | Glass | Chamber | 359.7 | 14.9 | 1.4 % |
|  |  | Coverslip | 364.7 | 17.4 |  |
|  | Graphene | Chamber | 419.6 | 22.0 | 6.4 % |
|  |  | Coverslip | 394.4 | 19.9 |  |

**Supplementary results on atomic force microscopy (AFM) characterization of graphene**

We performed AFM analyses over large areas of the graphene samples adopted in this study to investigate the graphene topography. In Figure S2 a representative AFM micrograph of 10 μm x 10 μm and relative line profile are reported. The root mean square (rms) roughness is found to be 1.8 nm, comparable to that of the underlying glass substrate [1].

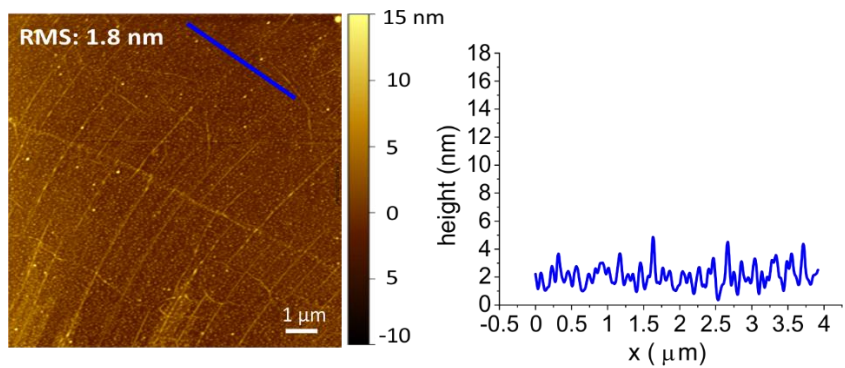

*Figure S2.* *10 μm x 10 μm AFM micrograph of transferred graphene and relative AFM height profile measured along the blue line.*

**Supplementary results on retrograde transport of vesicles containing Nerve Growth Factor**

*Table S3. Percentage of vesicles for each movement category at DIV2 and DIV5.* *Average percentage ± s.e.m. of vesicles in the time-lapse images of three independent cultures represented in the box plot in Figure 2e in the main text.*

| Movement categories | Substrate | Percentalge | s.e.m.[1] | DIV |
|---|---|---|---|---|
| Stalled | Graphene | 86.9 | 1.4 | DIV2 |
| | Control | 76.9 | 1.3 | |
| | Graphene | 70.3 | 2.0 | DIV5 |
| | Control | 68.4 | 1.1 | |
| Retrograde | Graphene | 1.0 | 0.2 | DIV2 |
| | Control | 8.5 | 1.0 | |
| | Graphene | 12.1 | 1.1 | DIV5 |
| | Control | 16.0 | 1.6 | |
| Anterograde | Graphene | 1.2 | 0.2 | DIV2 |
| | Control | 1.3 | 0.2 | |
| | Graphene | 1.6 | 0.2 | DIV5 |
| | Control | 1.0 | 0.1 | |
| Oscillating | Graphene | 9.7 | 0.8 | DIV2 |
| | Control | 12.2 | 0.6 | |
| | Graphene | 12.8 | 0.7 | DIV5 |
| | Control | 11.4 | 0.7 | |

[1] *Standard error of the mean (s.e.m.) of the percentage of vesicles for each population (stalled, retrograde, anterograde and oscillating) at DIV2 and DIV5. The number of acquired time-lapse images was 81 for glass and 56 for graphene at DIV2 and 75 for glass and 72 for graphene at DIV5. The number of vesicles n in three independent cultures is: DIV2, glass n=5188, graphene n=3604; DIV5, glass n=5399, graphene n=4905.*

**Supplementary results on measurements of carrier concentration in graphene**

Carrier concentration measurements were performed at room temperature using an Ecopia HMS-3000 Hall System operating in van der Pauw configuration, with the four golden probes placed in the corner of 5 x 5 mm substrates. Table S4 shows the carrier concentration of two representative samples of pristine graphene, after coating and after overnight in cell medium. In all the presented cases, the positive values displayed for the sheet concentration confirm holes as majority carriers in all the conditions, demonstrating an intrinsic p-type doping of the graphene substrates.

*TableS4. Sheet carrier concentration for pristine graphene, after coating and after cell medium.*

| Sample | | Sheet concentration (cm²) |
|---|---|---|
| 1 | As transfer | $2.11 \times 10^{13}$ |
| | Post coating | $1.6 \times 10^{13}$ |
| | Post medium | $9.99 \times 10^{12}$ |
| 2 | As transfer | $2.29 \times 10^{13}$ |
| | Post coating | $1.22 \times 10^{13}$ |
| | Post medium | $6.75 \times 10^{12}$ |

**Supplementary results on Raman characterization of graphene**

An important benchmark for the quality of graphene is the FWHM of the 2D mode. The broadening of the 2D peak is attributed to nanoscale strain fluctuations>. Figure S3b shows the map of the 2D-mode FWHM below or close to the neuron in Figure S3a, the same neuron-graphene system reported in Figure 3 in the main text, revealing that the 2D mode gets narrower for graphene under the neuron. This is supported by the 2D FWHM distribution, reported in Figure S3c; in the bimodal distribution the peak close to 30 cm$^{-1}$ is mainly due to the area under the cell meanwhile the peak at 37 cm$^{-1}$ is from "bare graphene" areas. However, this sharpening effect of the 2D peak is not homogeneous under the cell: there is an area on the right side of the cell where the 2D FWHM shows an increase up to 40 cm$^{-1}$, demonstrating that the interaction between graphene and neurons has a complex behavior.

Figures S3d and e show the maps of the energies of 2D and G modes in graphene, obtained on the same neuron/graphene system. In both maps, the G and 2D modes are shifted to higher Raman shifts where the neuron interact with the graphene. The position of the G and 2D modes is employed as an indicator for possible tensile/compressive strain of the graphene lattice >. The correlation plot in panel f shows that the area under the cell suffers a tensile stress in comparison to the bare graphene [2]. This tensile stress can be due to the growth of neurites after 2 days from cell deposition.

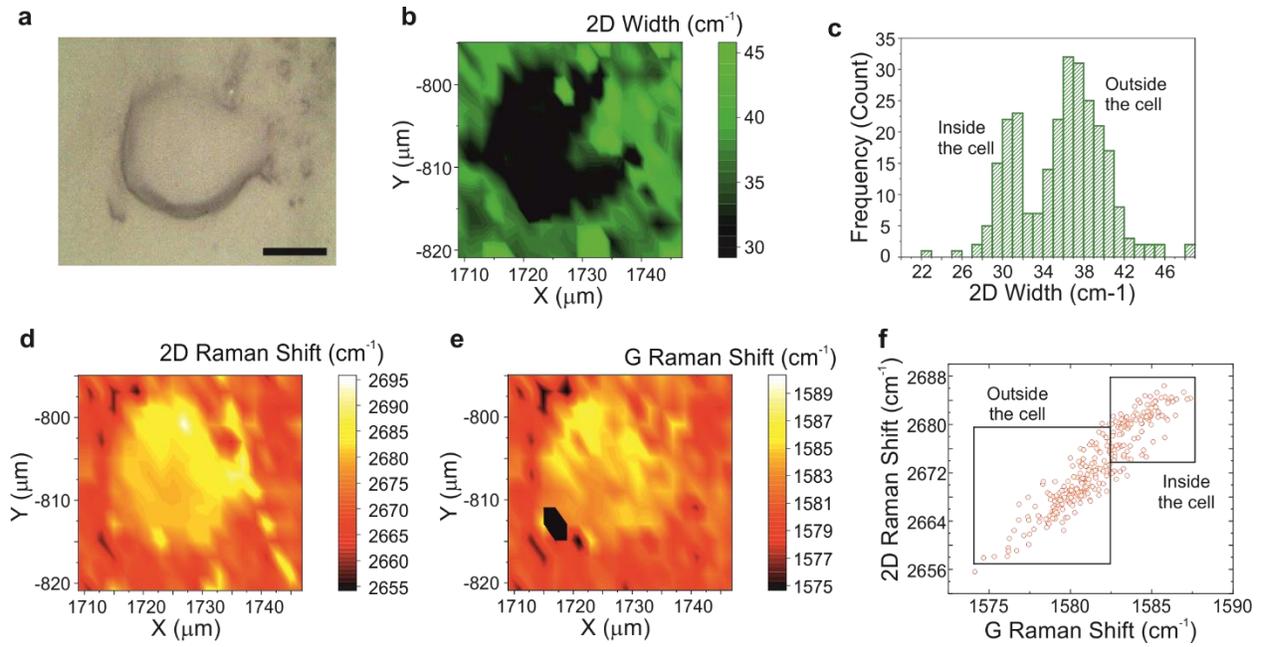

*Figure S3. Raman peaks characterization of the graphene-DRG neuron interface* (a) Bright field of the DRG neuron on graphene reported in Figure 3 in the main text. (b) Raman map of the 2D mode FWHM of the same area. (c) Histogram of 2D mode FWHM showing its bimodal distribution due to graphene outside or underneath the neuron. (d) Raman map of the position of the 2D mode. (e) Raman map of the position of the G mode. (f) Correlation plot of the 2D-G Raman peaks position.

Figure S4a shows the 2D/G intensity ratio distribution of a bare graphene area after the transfer, after coating with PDL/laminin and after the cell growth. There are no big differences amongst the distributions, indicating at most a homogeneous graphene doping upon coating and cell growth. In case of pristine graphene, the distribution has the peak with relative FWHM at (4.5 ± 1.2), after the coating the distribution peak increases up to (5±1) with a higher tail at low values. After the cell growth the distribution has the peak at (4.5 ± 1.3).

Panel (b) shows the 2D FWHM distribution for pristine graphene after the transfer, after coating and after cell growth. In all the presented cases, the distribution is peaked at 27 cm$^{-1}$ with a FWHM of 2 cm$^{-1}$ and higher tail at higher values, demonstrating the presence of areas with larger strain fluctuations.

We observed a discrepancy between the values of 2D/G intensity and 2D width after cell growth reported in panel (a) and (b) of Figure S4 and the values reported in figure 3g and figure S3c respectively. We ascribe this difference to sample fixation.

It is worth noting that the defect activated Raman mode (D mode) [4,>] is not present in all the conditions, indicating the high quality of the employed material and that the PDL/laminin coating and cell growth do not induce any damaging of the underlying graphene.

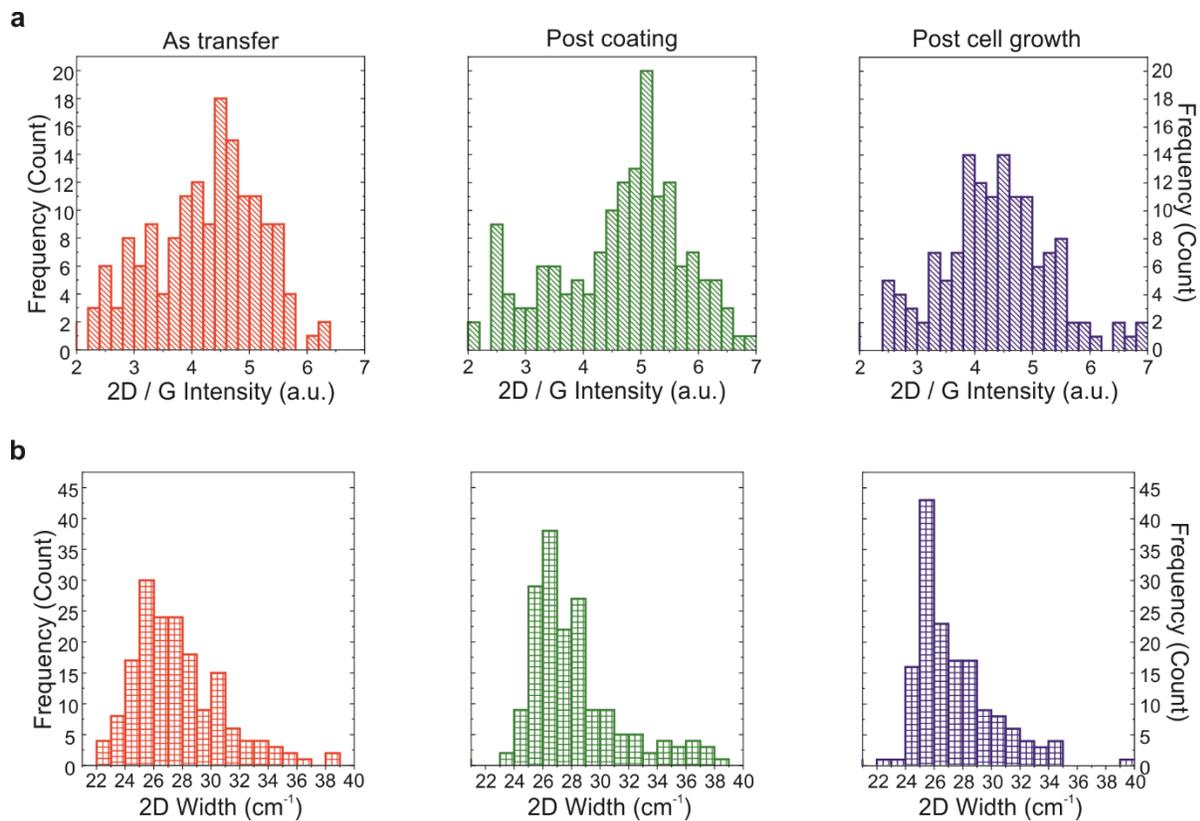

*Figure S4. Raman 2D/G intensity ratio distribution* *(a) 2D/G intensity ratio distribution for bare graphene (red), graphene after coating with PDL/laminin (green) and after the cell growth (blue). (b) 2D FWHM distribution for bare graphene (red), graphene after coating with PDL/laminin (green) and after the cell growth (blue).*

**Supplementary results on TEM analysis of microtubule organization**

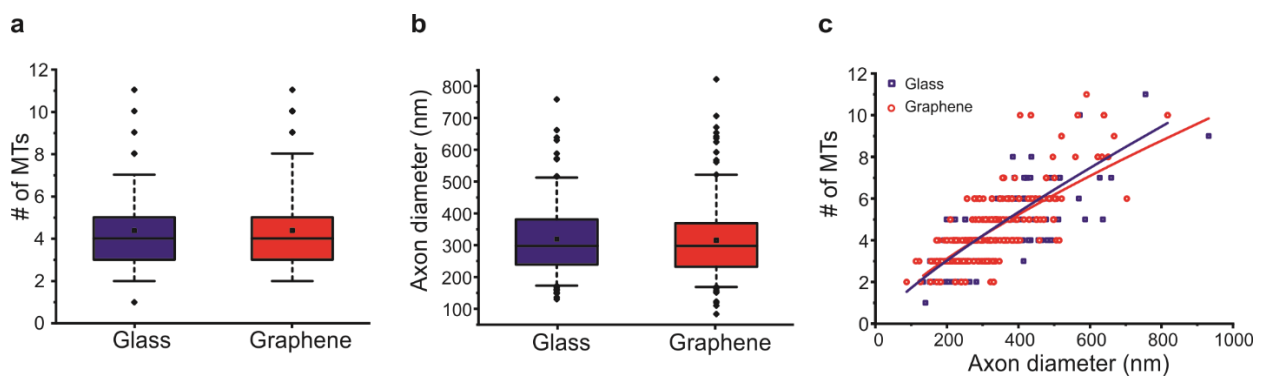

***Figure S5. TEM analysis of microtubule (MT) organization***: *(a,b) Box plots for number of MTs per axon and axon diameter for cells grown on graphene or glass. The differences are not significant. In the box plots: box between 25th and 75th percentile; horizontal line: median; whiskers: 5$^{th}$ and 95$^{th}$ percentiles; square: mean; circles: outliers. The number of analyzed axons from two independent culture was 79 for graphene and 97 for glass. (c) Dependency of the number of microtubules and axon diameter for cells grown on graphene (red circle) and glass (blue circle). The populations overlap and are not distinguishable. The number of MT ($N_{MT}$) dependency on axon diameter (D) can be fitted by the power law previously reported >*

$$N_{MT} = aD^m$$

*where a and m are the fitting parameters for the power law, with a=0.0595 and m=0.7472 for glass and a=0.0390 and m=0.8215 for graphene.*

**Supplementary results on DRG tubulin quantification**

In order to compare tubulin content in the axons of DRG neurons on graphene and control, we quantified the intensity of fluorescence in fixed DRG neurons stained with anti-βIII-tubulin antibodies and Alexa488 secondary antibodies. We selected ROIs containing axons, removing contributions of cell bodies and glial cells (clearly visible thank to a phalloidin staining), and we calculated the mean tubulin intensity by averaging the mean intensity inside each ROI.

We observed that in the case of graphene the axons were on the same focal plane of glial cells closer to the substrate, probably due to an increased neural adhesion.

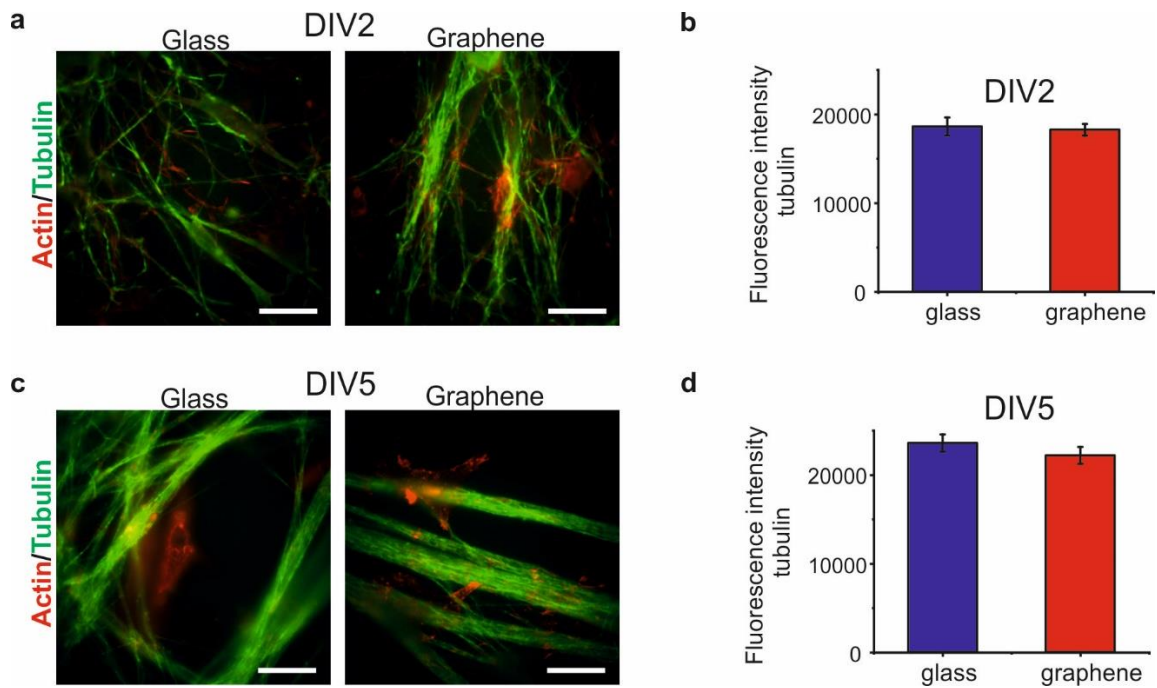

*Figure S6. DRG neurons tubulin quantification (a,c) Representative fluorescence microscopy images of neural culture stained with phalloidin to identify actin and β-III tubulin to identify axons at DIV2 (a) and DIV5 (c). (b.d) Mean ± s.e.m. of the fluorescence intensity of Alexa488-labelled tubulin in DRG neurons in the selected ROI at DIV2 (b) and DIV5 (d). More than 25 fields were analyzed for each condition. The number of acquired ROI containing axons was 31 for glass and 51 for graphene at DIV2 and 31 for glass and 28 for graphene at DIV5.*